\documentclass[sort&compress,4apaper]{aipproc}
\layoutstyle{8x11single}
\usepackage{graphicx,psfrag}
\usepackage{epsfig}
\usepackage{amsmath}
 \usepackage{amscd}
 \usepackage{hyperref}
 \usepackage{amssymb}
\usepackage{epsf,float}
\usepackage{bm}%

\begin{document}
 \title{Clockwork Quantum Universe\\ \small{IV \textit{Prize, } FQXi Essay Contest 2011: \textit{  ``Is Reality Digital or Analog?''}}}

\author{Donatello Dolce}{
address={The University of Melbourne, School of Physics, Parkville VIC 3010, Australia.},
,email= {ddolce@unimelb.edu.au}}

\begin{abstract}
Besides the purely digital or analog interpretation of reality there
is a third possibility which incorporates important aspects of both.
This is the cyclic formulation of elementary systems, in which 
 elementary particles are represented as classical strings vibrating in compact
space-time dimensions with periodic boundary conditions. We will address these cyclic solutions as ``de
Broglie internal clocks''. They constitute the deterministic gears
of a consistent semi-classical description of quantum relativistic
physics, providing in addition an appealing formulation of the notion
of time. 
\end{abstract}
\maketitle

\section{Introduction}

One of the possible ways to introduce the cyclic interpretation of elementary particle physics \cite{Dolce:2009ce,Dolce:tune,Dolce:AdSCFT,Dolce:2009cev4,Dolce:Dice,Dolce:2010ij,Dolce:2010zz,Dolce:cyclic,Dolce:SuperC}
 is provided by \text{'t Hooft's} 
determinism \cite{'tHooft:2001ar,'tHooft:2001fb,Elze:2002eg,'tHooft:2006sy}. It
states that ``there is a close relationship between a quantum harmonic
oscillator'' with angular frequency $\bar{\omega}=2\pi/T_{t}$, \emph{e.g.} a single mode of an ordinary
second-quantized field of energy $\bar{E}=\hbar\bar{\omega}$, ``and a classical particle moving along a circle'' with time periodicity
$T_{t}$. By assuming the time period $T_{t}$ on a lattice with $N$
sites, it turns out that if the experimental time accuracy is too low ($\Delta t\gg T_{t}$),
at every observation the system appears in an arbitrary discretized phase of its
cyclic evolution, \textit{i.e} on an arbitrary site of the periodic lattice.
Since the underlying periodic dynamics are too fast to be observed,
the evolution has an apparent aleatoric behavior as if observing a
``clock under a stroboscopic light'' \cite{Elze:2002eg}. The
evolution operator $\mathcal{U}(\Delta t=\epsilon)=\exp[{-\frac{i}{\hbar}\mathcal{H}\epsilon}]$
is given in terms of a $N\times N$ matrix and the model is analogous
to a harmonic system of $N$ masses and springs on a ring. In the
limit of large $N$, the frequency eigenvectors $|\phi_{n}\rangle$
obey to the relation $\mathcal{H}|\phi_{n}\rangle\sim\hbar\bar{\omega}\left(n+{1}/{2}\right)|\phi_{n}\rangle$
which actually describes the energy eigenvalues $E_{n+1/2}=\hbar\bar{\omega}\left(n+{1}/{2}\right)$
of a quantum harmonic oscillator with periodicity $T_{t}$ - apart
for an ``unimportant'' phase in front of the operator $U(\epsilon)$
which reproduces the factor $1/2$ in the eigenvalues \cite{'tHooft:2006sy,'tHooft:2007xi,Jaffe:2005vp} 
and which can be regarded a twist factor on the Periodic Boundary Conditions
(PBCs). The idea is that, due to the extremely fast cyclic dynamics,
we loose information about the underlying classical theory and we
observe a statistical theory that matches QM. For this reason we speak
about deterministic or pre-quantum theories. Since the cyclic time
interval $T_{t}$ is supposed on a lattice, the t' Hooft determinism
can be classified as purely digital. It is recently evolved into 
the idea of ``classical cellular automata'' \cite{'tHooft:2010zzb}  (\emph{i.e.} a deterministic model
with interesting correspondences between elementary particles
and black holes). However, if we take the continuous limit of the t'
Hooft deterministic model by assuming an infinite number of lattice sites $N\rightarrow\infty$,
it is easy to see that the system of springs and masses turns out to be a
vibrating string embedded in a cyclic time dimension, that is a bosonic
classical field $\Phi(x,t)$ embedded in a compact time dimension of length
$T_{t}$ and PBCs. Formally, through discrete Fourier transform,
to a compact variable corresponds a quantized conjugate variable,
that is to say a variable which takes discrete values. Hence, a compact dimension yields to a digital description
of the conjugate space. Considering the 
relation $\bar{E}=\hbar\bar{\omega}$, to the intrinsically periodic system with
 $t\in[0,T_{t}]$ there is associated  the quantized energy spectrum
$E_{n}=n\hbar\bar{\omega}=nh/T_{t}$. The energy is the digital conjugate
variale of a cyclic time variable. More in general, since we experimentally observe
an energy-momentum space on a lattice (quantized energy-momentum spectrum), it is natural to try to describe QM in terms
of intrinsic space-time periodicities. In \cite{Dolce:2009ce,Dolce:tune} we have shown
that, similarly to a particle in a box, relativistic fields can be actually
quantized by imposing their characteristic de Broglie space-time  periodicities
as constraints. 

Our assumption of dynamical periodic fields can be regarded as a combination
of the Newton's law of inertia and de Broglie hypothesis of
undulatory mechanics: \textit{elementary isolated systems must be
supposed to have persistent periodicities as long as they do not interact}.
Such an assumption of intrinsic periodicity is also implicit in the
operative definition of time (and for some aspects in the action-reaction law). Time can only be defined by counting
the number of cycles of isolated phenomena \textit{supposed} to be
periodic. For a consistent formalization  of time in physics,  there must be  an assumption of intrinsic periodicity
for free elementary systems!  In modern physics
a second is defined as the duration of 9,192,631,770 characteristic  \emph{cycles}
of the Cs atom ($T_{Cs}\sim10^{-10}s$). For the central role of time in physics, the assumption
of isochronism of the pendulum made by Galileo in the cathedral of
Pisa can be regarded as one of the foundational acts of physics. Such
an assumption of persistent periodic phenomena allowed a sufficiently
accurate definition of time to study  the motion of bodies and
in turn the formulation of theories of dynamics. The definition
of relativistic clock given by A. Einstein \cite{Einstein:1910} is: ``by a
clock we understand anything characterized by a phenomenon passing
periodically through identical phases so that we must assume, by the
principle of sufficient reason, that all that happens in a given period
is identical with all that happens in an arbitrary period''. The
whole information of such a relativistic clock is contained in a single
 period. Thus, by using the terminology of extra dimensional theories, in reference clocks time can be formalized as a compact, analog dimension with PBCs.  In this way   every free particle, represented as a non-interacting cyclic field with intrinsic de Broglie time periodicities $T_{t}$, can be
regarded of as reference clock, also known as  ``de Broglie internal clocks''
\cite{1996FoPhL,2008FoPh...38..659C}. 

Since the measure of time is a counting
process it also has a digital nature. This intrinsically leads to
the Heisenberg uncertain principle, \cite{Dolce:2009ce,Dolce:2009cev4}. In fact, in a `` de Broglie clock'', to determine the energy
$\bar{E}=\hbar\bar{\omega}$ with good accuracy $\Delta\bar{E}$ we
must count a large number of cycles, that is to say we must observe the system
for a long time $\Delta t$, according to the relation $\Delta\bar{E}\Delta t\gtrsim\hbar$. 
Moreover, since intrinsic periodicity means that the only possible
energy eigenmodes are those with an integer number of cycles, we obtain
the Bohr-Sommerfeld quantization condition (it can be shown that the periodicity
condition $E_{n}T_{t}=nh$ can be more in general written as $\oint E_{n}dt=nh$ for interacting systems).
Intrinsic periodicity can be in fact used to solve non-relativistic quantum problems \cite{Dolce:2009ce,Dolce:2009cev4}.

 For a Lorentz covariant formulation of the theory we must consider that the de Broglie time periodicity  induces spacial de Broglie periodicities
$\lambda^{i}$, and that these space-time periodicities, as well as
the energy-momentum quantized spectrum, transforms in a relativistic way. In other words, since $T_{t}=h/\bar{E}$, the de Broglie time periodicity
must be regarded as dynamical. As every time interval and in analogy with the Doppler effect,
$T_{t}$ transforms in a relativistic way. The proper-time intrinsic
periodicity $T_{\tau}$ fixes the upper bond of the time periodicity
$T_{t}$ because the mass is the lower bond of the energy. For instance, 
by denoting the reference system by the spatial momentum  $\mathbf{\bar{p}}$,
where $p_{i}=h/\lambda^{i}$, we have $T_{\tau}\geq T_{t}(\mathbf{\bar{p}})$
and $\bar{M}c^{2}\leq\bar{E}(\mathbf{\bar{p}})$. The heavier the
mass the faster the proper-time periodicity. Hence, even a light particle
such as the electron has (in a generic reference frame) intrinsic
time periodicity equal or faster than $\sim10^{-20}s$, \textit{i.e.}
the time periodicity in a generic reference frame is always faster than its proper-time periodicity. It should be noted that such a periodicity
is many orders of magnitude away from the characteristic time periodicity
of the cesium atomic clock, which by definition is of the order of
$10^{-10}s$, and that it is extremely fast even if compared with the present
experimental resolution in time ($\sim10^{-17}s$). Thus, for every
known  matter particle (with the exception of neutrinos) we are in the case of too
fast periodic dynamics as in the 't Hooft determinism. The de Broglie
intrinsic clock of elementary particles can also be imagined as a dice, 
named ``de Broglie deterministic dice'' \cite{Dolce:Dice},  rolling with
time periodicity $T_{t}$. In fact we inevitably have a too low revolution
in time, so that at every observation the system appears in an aleatoric
phase of its evolution. Indeed, as for a clock under a stroboscopic
light or a dice rolling too fast with respect to our time resolution, we can only predict the outcomes
statistically (an observer with infinite time resolution would not have fun playing dices). From the results presented in \cite{Dolce:2009ce,Dolce:tune} and
summarized here we will see that such a statistical description associated
to intrinsically periodic phenomena formally matches ordinary QM.
We may also note that, on a cyclic geometry such as that associated with intrinsic periodicity, there exist many possible classical paths, characterized by different winding (digital) numbers,
between every initial and final point. Thus the evolution of a cyclic field is described by a sum over classical cyclic paths. The result is a formal matching with the ordinary Feynman Path Integral. 
In this essay we will only describe some published results or announce
some others that will be published soon.  The reader interested in more technical
details or to the mathematic proofs may refers to \cite{Dolce:2009ce,Dolce:tune,Dolce:2009cev4,Dolce:AdSCFT}.

\section{Relativistic gears}

The relativistic generalization of Newton's law of inertia can be stated as follows:
 every isolated elementary system has persistent four-momentum
$\bar{p}_{\mu}=\{\bar{E}/c,\mathbf{\bar{p}}\}$. On the other hand,
the de Broglie formulation of QM prescribes that to the four-momentum
must be associated a ``periodic phenomenon'' of four-angular-frequency, according to the relation $\bar{\omega}_{\mu}=\bar{p}_{\mu}c/\hbar$. 
Here we will assume that every elementary
system is described as field of intrinsic de Broglie periodicity $T^{\mu}=\{T_{t},\vec{\lambda}_{x}/c\}=2 \pi / \bar \omega_\mu$ imposed as constraint.
As the Newton's law of inertia does not imply that every point particle
moves on a straight line, our assumption of intrinsic periodicities
does not mean that the physical world should appear to be periodic.
In fact, the four-periodicity $T^{\mu}$ is fixed dynamically by the
four-momentum through the de Broglie-Planck relation 
\begin{equation}
T^{\mu} \bar{p}_{\mu}c =h \,.\label{eq:PdB:relation}
\end{equation}
From this  follows that the variation of four-momentum occurring during interactions
implies a corresponding modulation of the intrinsic periodicity of the fields.
This guarantees time ordering and relativistic causality. 

Similarly to the 't Hooft deterministic  model, a free cyclic field $\Phi(\mathbf{x},t)$
is a tower of frequency eigenmodes $\phi_n(x)$ with energies $E_{n}(\bar{\mathbf{p}})=n \bar {E}(\mathbf{\bar{p}})=n\hbar\bar{\omega}(\mathbf{\bar{p}})$,
\begin{equation}
\Phi(\mathbf{x},t)=\sum_{n}A_{n}\phi_{n}(\mathbf{x})u_{n}(t)~,~~~~~~~~\text{where}~~~~~~u_{n}(t)=e^{-i\omega_{n}(\mathbf{\bar{p}})t}\,.\label{field:exp:modes}\end{equation}
By bearing in mind the relation $\bar{E}(\bar{\mathbf{p}})=\hbar\bar{\omega}(\mathbf{\bar{p}})$,
the quantized energy spectrum $E_{n}(\bar{\mathbf{p}})$ is nothing
but the harmonic frequency spectrum ${\omega_{n}}(\mathbf{\bar{p}})=n\bar{\omega}(\mathbf{\bar{p}})$
of a string vibrating with time periodicity $T_{t}(\bar{\mathbf{p}})$%
\footnote{The theory can be regarded as a particular kind of string theory in which  the compact world-line parameter plays the role of the compact
world-sheet parameter.%
}. This quantization is the field theory analogous  of  the semiclassical
quantization of a ``particle'' in a box, it also shares deep analogies
with the Matsubara and the Kaluza-Klein (KK) theory \cite{Matsubara:1955ws}.
Since in this case the whole physical information of the system is
contained in a single four-period $T^{\mu}$, our intrinsically four-periodic
free field can be described by a bosonic action in compact space-time dimensions
with PBCs \begin{equation}
\mathcal{S}_{\lambda_{s}}=\int_{0}^{T^{\mu}}d^{4}x\mathcal{L}_{\lambda_{s}}(\partial_{\mu}\Phi,\Phi)\,.\label{free:act}\end{equation}
 It is important to note that PBCs minimize the action at the boundaries ---
in particular the ones of the time dimension. Therefore PBCs have the
same formal validity of the usual (Synchronous) BCs assumed in ordinary
field theory. In this way
all the symmetries of the relativistic bosonic theory are preserved as usual. In particular it guarantees that the theory is
Lorentz invariant. In fact we may consider a generic global Lorentz
transformation \begin{equation}
dx^{\mu}\rightarrow dx'^{\mu}=\Lambda_{\nu}^{\mu}~dx^{\nu}~,~~~~~~~~~~\bar{p}_{\mu}\rightarrow\bar{p}'_{\mu}=\Lambda_{\mu}^{\nu}~\bar{p}_{\nu}\,.\label{space:mom:Lorentz:tranf}\end{equation}
 The phase of the field
is invariant under four-periodic $T^{\mu}$ translations $\exp{[-ix^{\mu}\bar{p}_{\mu}]}=\exp{[-i(x^{\mu}+c T^{\mu})\bar{p}_{\mu}]}\,.$ and it is a scalar quantity under Lorentz transformations - de Broglie phase harmony.
In this way we find that, as every generic space-time interval, the four-periodicity is actually a contravariant four-vector \begin{equation}
T^{\mu}\rightarrow T'^{\mu}=\Lambda_{\nu}^{\mu}~T^{\nu}\,.\label{period:Lorentz:tranf}\end{equation}
 The four periodicity $T^{\mu}$
can be thought of as describing a reciprocal energy-momentum lattice
$p_{n\mu}=n\bar{p}_{\mu}$. Invariance can also be  inferred by noticing
that after the transformation of variables (\ref{space:mom:Lorentz:tranf}),
the integration region of the free action (\ref{free:act}) turns
out to be transformed as well, \begin{equation}
\mathcal{S}_{\lambda_{s}}=\int_{0}^{T'^{\mu}}d^{4}x'\mathcal{L}_{\lambda_{s}}(\partial'_{\mu}\Phi',\Phi')\,.\label{Lorentz:trans:action}\end{equation}
 Therefore, in the new reference system, the new four-periodicity
$T'^{\mu}$ of the field is actually given by (\ref{period:Lorentz:tranf}).
That is (\ref{Lorentz:trans:action}) describes a system with four-momentum
$\bar{p}'_{\mu}$ (\ref{space:mom:Lorentz:tranf}).

The underlying Minkowski metric induces the following  constraint on the dynamical  four-periodicities \[
\frac{1}{T_{\tau}^{2}}\equiv\frac{1}{T_{\mu}}\frac{1}{T^{\mu}}\]
 which, considering the above de Broglie-Planck relation, is nothing
but the relativistic constraint  $\bar{M}^{2}c^{2}=\bar{p}^{\mu}\bar{p}_{\mu}$.

The resulting compact 4D formulation reproduces, after normal ordering,
exactly the same quantized energy spectrum of ordinary second quantized
fields. In fact,  a cyclic field with mass $\bar{M}$ turns out to have energy spectrum \[
{E}_{n}(\mathbf{\bar{p}})=n \bar {E}(\mathbf{\bar{p}})=n\sqrt{\bar{\mathbf{p}}^{2}c^{2}+\bar{M}^{2}c^{4}}\]
 of ordinary quantum field theory. Furthermore, it is easy to see
that in the rest frame ($\mathbf{\bar{p}}\equiv0$) this quantized
energy spectrum is \textit{dual} to the KK mass tower
$M_{n}=E_{n}(0)/c^{2}=n\bar{M}$. Indeed, for such a massive field,
the assumption of periodicity along the time dimension means that
in the rest frame the proper-time $\tau$ there has intrinsic periodicity
 \[
T_{\tau}=T_{t}(0)=\frac{h}{\bar{M}c^{2}}\]
The invariant mass $\bar{M}$ is not a parameter of the Lagrangian. It is
 fixed geometrically by the reciprocal of the proper-time intrinsic
periodicity $T_{\tau}$ thought PBCs, and thus by the compactification lengths of the theory. In other words, by
imposing intrinsic time periodicity, the world-line parameter $s=c\tau$
turns out to be compact with PBCs. It behaves similarly to the XD
of a KK field with zero 5D mass and with fundamental mass $\bar{M}$.
As a consequence the world-line compactification length $\lambda_{s}=cT_{\tau}$
is the Compton wavelength of the field. In order to bear in mind
these analogies with an XD field theory we say that the world-line
parameter plays the role of a \emph{Virtual} XD (VXD) with compactification
length $\lambda_{s}$, \cite{Dolce:AdSCFT}. It is interesting to note that, originally,
T. Kaluza introduced the XD formalism as a ``mathematical trick''
and not as a \emph{real} XD \cite{Kaluza:1921tu}.

\section{Quantum gears}

Here we  show that our cyclic description of elementary particles
has a remarkable formal matching to the canonical (axiomatic) formulation of QM as well
as to the Feynman Path Integral (FPI) formulation. The evolution
along the compact time dimension is described by the so called bulk
equations of motion $(\partial_{t}^{2}+\omega_{n}^{2})\phi_{n}(x,t)=0$
- for the sake of simplicity in this section we assume a single spatial
dimension $x$. 
Thus the time evolution of the energy eigenmodes can be written as
first order differential equations $i\hbar\partial_{t}\phi_{n}({x},t)=E_{n}\phi_{n}({x},t)$.
 The periodic field (\ref{field:exp:modes}) is a sum of on-shell
standing waves. Actually this is the typical case where a Hilbert
space can be defined. In fact, the energy eigenmodes form a complete
set with respect to the inner product \begin{equation}
\left\langle \phi|\chi\right\rangle \equiv\int_{0}^{\lambda_{x}}\frac{{dx}}{{\lambda_{x}}}\phi^{*}(x)\chi(x)\,.\label{inner:prod}\end{equation}
 Therefore the energy eigenmodes  define  Hilbert eigenstates
$\left\langle {x}|\phi_{n}\right\rangle \equiv{\phi_{n}({x})}/{\sqrt{\lambda_{x}}}$.
On this base we can formally build a Hamiltonian operator $\mathcal{H}\left|\phi_{n}\right\rangle \equiv\hbar\omega_{n}\left|\phi_{n}\right\rangle $
and a momentum operator $\mathcal{P}\left|\phi_{n}\right\rangle \equiv-\hbar k_{n}\left|\phi_{n}\right\rangle $,
where $k_{n}=n\bar{k}=nh/\lambda_{x}$. Thus the time evolution of
a generic state $|\phi(0)\rangle\equiv\sum_{n}a_{n}|\phi_{n}\rangle$
is described by the familiar Schr\"odinger equation \begin{equation}
i\hbar\partial_{t}|\phi(t)\rangle=\mathcal{H}|\phi(t)\rangle.\end{equation}
 Moreover the time evolution is given by the usual time evolution
operator $\mathcal{U}(t';t)=\exp[{-\frac{{i}}{\hbar}\mathcal{H}(t-t')}]$
which turns out to be a Marcovian (unitary) operator: $\mathcal{U}(t'';t')=\prod_{m=0}^{N-1}\mathcal{U}(t'+t_{m+1};t'+t_{m}-\epsilon)$
where $N\epsilon=t''-t'~$.

From the fact that the spatial coordinate is in this theory a cyclic
variable; by using the definition of the expectation value of an observable
$\hbar\partial_{x}F(x)$ between two generic initial and final states
$|\phi_{i}\rangle$ and $|\phi_{f}\rangle$ of this Hilbert space;
and integrating by parts (\ref{inner:prod}), we find \begin{equation}
\left\langle \phi_{f}|\hbar\partial_{x}F(x)|\phi_{i}\right\rangle = i \left\langle \phi_{f}|\mathcal{P}F(x)-F(x)\mathcal{P}|\phi_{i}\right\rangle \,.\end{equation}
 By assuming that the observable is such that $F(x)=x$ \cite{Feynman:1942us}
we obtain the usual commutation relation of ordinary QM: $[x,\mathcal{P}]=i\hbar$  --- or more in general $[F(x),\mathcal{P}]=i\hbar \partial_x F(x)$. The commutations relations are implicit in the assumption of intrinsic periodicity.
With this result we have checked the correspondence with canonical
QM. 

Similarly, it is possible to prove the correspondence with the
FPI formulation. In fact, it is sufficient to plug the completeness
relation of the energy eigenmodes in between the elementary time evolutions
of the Marcovian operator.  With this elements at hand and proceeding
in a complete standard way we find that the evolution of the cyclic fields turns out to be described by the usual FPI which, in phase space ($V_x = N \lambda_x$ with $N\in \mathbb N$  large in case of interaction, see \cite{Dolce:2009ce,Dolce:tune}),
can be written as \begin{equation}
\mathcal{Z}=\lim_{N\rightarrow\infty}\int_{0}^{V_{x}}\left(\prod_{m=1}^{N-1}\frac{dx_{m}}{V_x}\right)\prod_{m=0}^{N-1}\left\langle \phi\right|e^{-\frac{i}{\hbar}(\mathcal{H}\Delta\epsilon_{m}-\mathcal{P}\Delta x_{m})}\left|\phi\right\rangle \,,\label{periodic:path.integr:Oper:Fey}\end{equation}

This important result has been obtained without any further assumption
than PBCs and has a simple classical interpretation. In a cyclic geometry
there is an infinite set of possible classical paths with different
winding numbers that link every given initial and final points. If
we imagine to open this cyclic geometry we obtain a lattice with period
$T^{\mu}$ of initial and final points linked by classical paths.
The FPI obtained in (\ref{periodic:path.integr:Oper:Fey}) means that a cyclic field can self-interfere and its classical evolution is described by summing over the possible paths with different winding number. These path play the role of the non-classical paths of Feynman interpretation of QM, though the  are classical paths minimizing the action in compact space-time dimensions. This means
that in this path integral formulation it is not necessary to relax
the classical variational principle  to have self-interference.

The non-quantum limit of a massive field, \emph{i.e.} the non-relativistic
single particle description, is obtained by putting the mass to infinity
so that, as shown in \cite{Dolce:2009ce,Dolce:2009cev4}, in such an effective
classical limit, only the first level of the energy spectrum must
be considered, \cite{Dolce:2009ce,Dolce:2009cev4}. This leads to a consistent interpretation of the wave-particle
duality and of the double slit experiment. The quantities describing
only the first energy level are addressed by the bar sign. For instance,
the Lagrangian of the transformed fundamental mode $\bar{\Phi}'(x')$ is $\mathcal{\bar{L}}_{\lambda_{s}}(\partial_{\mu}\bar{\Phi}'(x'),\bar{\Phi}'(x'))$.
Note that the transformed fundamental mode $\bar \Phi'(x')$ coincides with the mode of Klein-Gordon
field with energy $\bar{E}'$ and mass $\bar{\mbox{M}}$. Therefore it can be always  quantized through second quantization. For this reason the analysis of the geometrodynamics of the de Broglie periodicities that we will perform below can be extended to ordinary field theory, \cite{Dolce:tune}. On the other hand a massless field has infinite Compton wavelength and thus an infinite proper-time periodicity. Its quantum limit is at high frequency. In this limit the PBCs are important and we have a discretized energy spectrum, in agreement with the ordinary description of the black-body radiation (no UV catastrophe). The opposite limit described by a continuous energy spectrum is when time periodicity tends to infinity. 

In the original 't Hooft model the period $T_{t}$ was assumed to be of the
order of the Planck time, in an attempt to avoid hidden variables, \cite{Nikolic:2006az}. Furthermore the Hamiltonian operator was not positive defined. In our case, the intrinsic periodicities are reproduced by compact dimensions with PBCs. Therefore we have the remarkable property that QM emerges without involving any hidden-variable. The theory can in principle violates the Bell's inequality and we can actually speak about determinism. Moreover, similarly to the KK theory in which there are no tachyons, a cyclic field can have positive or negative frequency eigenmodes but the energy spectrum describes always positive energies and the Hamiltonian operator is positive defined.

\section{Geometrodynamics}

To introduce interactions we must bear in mind that the four-periodicity
$T^{\mu}$ is dynamically related to the four-momentum $\bar{p}_{\mu}$
according to the de Broglie-Planck relation (\ref{eq:PdB:relation}).
An isolated elementary system (\textit{i.e.} free
field) has persistent four-momentum, whereas  an elementary
system under a generic interaction scheme can be described in terms
of corresponding variations of four-momentum along its evolution with
respect to the free case \begin{equation}
\bar{p}_{\mu}\rightarrow\bar{p}'_{\mu}(x)=e_{\mu}^{a}(x)\bar{p}_{a}\,.\label{eq:deform:4mom:generic:int}\end{equation}
The tetrad (or virebein) $e_{\mu}^{a}(x)$ turns out to encode the interaction scheme. But in undulatory mechanics the interaction 
(\ref{eq:deform:4mom:generic:int}) can be equivalently described in terms
of corresponding local modulations of  space-time periodicity \begin{equation}
T^{\mu}\rightarrow T'^{\mu}(x)\sim e_{a}^{\mu}(x)T^{a}\,.\label{eq:deform:4period:generic:int}\end{equation}
In our formalism  these modulations correspond to local deformations of the compactification lengths. Roughly speaking, interactions can be formalized
 as local stretching of the compact dimensions of the theory. Therefore
the interaction (\ref{eq:deform:4mom:generic:int}) turns out to
be equivalently encoded in a corresponding curved space-time background, which
in the limit of weak interaction can be approximated as \begin{equation}
\eta_{\mu\nu}\rightarrow g_{\mu\nu}(x)\sim e_{\mu}^{a}(x)e_{\nu}^{b}(x)\eta_{ab}\,.\label{eq:deform:metric:generic:int}\end{equation}

This result can be checked by considering the transformation
of space-time variables \begin{equation}
dx_{\mu}\rightarrow dx'_{\mu}(x)\sim e_{\mu}^{a}(x)dx_{a}\,.\label{eq:deform:mesure:generic:int}\end{equation}
 Under the approximation of weak interaction we are assuming that
the $T^{\mu}$ transforms as an infinitesimal interval $dx^{\mu}$.
After this transformation of variables (diffeomorphism) with determinant
of the Jacobian $\sqrt{-g(x)}$, the free action (\ref{free:act})
turns out to be\footnote{
For the sake of simplicity, we work in the approximation in which the compactification length can be identified with the de Broglie periodicity, \cite{Dolce:tune}. A more exact description would involve Christoffel symbols.} \begin{equation}
\mathcal{S}_{\lambda_{s}}\sim\int^{e_{a}^{\mu}(x)T^{a}}d^{4}x'\sqrt{-g}\mathcal{L}_{\lambda_{s}}(e_{\mu}^{a}\partial_{a}\Phi',\Phi')\,.\label{eq:defom:action:generic:int}\end{equation}
Indeed, the transformed periodic field $\Phi'(x')$ which minimizes this action has four-periodicity
$T'^{\mu}$, (\ref{eq:deform:4period:generic:int}), or equivalently
has four-momentum $\bar{p}_{\mu}$, (\ref{eq:deform:4mom:generic:int}).
We conclude that a field under the interaction scheme (\ref{eq:deform:4mom:generic:int})
is described by the solutions of the bulk equations of motion  on the deformed compact background (\ref{eq:deform:metric:generic:int})
and compactification lengths (\ref{eq:deform:4period:generic:int}).

 Our geometrodynamical approach to interactions is interesting because
it mimics very closely the usual geometrodynamical approach
of GR. In fact, it is important to note that gravitational interaction can be interpreted as modulations of periodicity of reference clocks. For instance we may consider a weak Newton potential $V(\mathbf{x})=-{GM_{\odot}}/{|\mathbf{x}|}\ll1$.
The energy on a gravitational well varies (with respect
to the free case) as $\bar{E}\rightarrow\bar{E}'\sim\left(1+{GM_{\odot}}/{|\mathbf{x}|}\right)\bar{E}$.
According to (\ref{eq:deform:4period:generic:int}) or (\ref{eq:PdB:relation}),
this means that the de Broglie clocks in a gravitational well are
slower with respect to the free clocks $T_{t}\rightarrow T_{t}'\sim\left(1-{GM_{\odot}}/{|\mathbf{x}|}\right)T_{t}$.
Thus we have a gravitational redshift $\bar{\omega}\rightarrow\bar{\omega}'\sim\left(1+{GM_{\odot}}/{|\mathbf{x}|}\right)\bar{\omega}$.
With our simple schematization of interactions we have retrieved two important
predictions of GR.
Besides this we must also consider the analogous variation of spatial
momentum and the corresponding modulation of spatial periodicities
\cite{Ohanian:1995uu}. According to (\ref{eq:deform:metric:generic:int})
the weak newtonian interaction turns out to be encoded in the usual linearized
Schwarzschild metric \begin{equation}
ds^{2}\sim\left(1-\frac{{GM_{\odot}}}{|\mathbf{x}|}\right)dt^{2}-\left(1+\frac{{GM_{\odot}}}{|\mathbf{x}|}\right)d|\mathbf{x}|^{2} - |\mathbf{x}|^2 d \Omega^2 \,.\end{equation}
 We have found that the geometrodynamical approach to interactions
actually can be used to describes linearized gravity and that the
geometrodynamics of the compact space-time dimensions correspond to
the usual relativistic ones.

As well known, see for instance \cite{Ohanian:1995uu}, it is possible
to retrieve ordinary GR from a linear formulation by including self-interactions.
More naively, as we will show in detail in  forthcoming papers, we may regard the metric  $g_{\mu\nu}$ a a dynamical field of the theory.  This corresponds to introduce ``by hand'' a kinetic term (with appropriate coupling $16\pi G_{N}$)
to the Lagrangian in curve space-time (\ref{eq:deform:4mom:generic:int}).
 Moreover, in order to neglect
quantum corrections, we may replace the Lagrangian $\sqrt{-g}\mathcal{L}_{\lambda_{s}}$
of (\ref{eq:defom:action:generic:int}) with its non-quantum limit
$\sqrt{-g}\mathcal{\bar{L}}_{\lambda_{s}}$ (\emph{i.e.} the Lagrangian of the fundamental mode $\Phi$). Thus we obtain the familiar 
 Hilbert-Einstein Lagrangian \begin{equation}
{\mathcal{\bar{L}}}_{HE}=\sqrt{-g}\left[-\frac{g^{\mu\nu}\mathcal{R}_{\mu\nu}}{16\pi G_{N}}+\mathcal{\bar{L}}_{\lambda_{s}}(e_{\mu}^{a}\partial_{a}\bar{\Phi}',\bar{\Phi}')\right]\,.\label{eq:defom:action:generic:int}\end{equation}
 This naive procedure is similar to the derivation of the kinematic term  $F_{\mu\nu}F^{\mu\nu}/-4e^{2}$ in gauge field theories \footnote{In \cite{Dolce:tune} we have also shown that it can be derived directly by applying the variational principle at the boundary of the theory.}.  Because of its geometrical
meaning, the Ricci tensor  is the correct mathematical object to describes
the dynamical variations of the space-time compactification lengths at different
interaction points. This can be intuitively seen by using the de Broglie harmony (\ref{eq:PdB:relation}) and that encodes the content of
four-momentum in different space-time points.  Here we also mention that it is not uniquely defined ``what is fixed at the boundary of the action principle of GR'', \cite{springerlink:10.1007/BF01889475}. As well known, Einstein's equation can be obtained from different
action formulations, differing by boundary terms.
In particular in the ordinary formulation of GR we can neglect issues related to the variations of the boundary terms and obtain  Einstein
equation $\mathcal{R}^{\mu\nu}=-8\pi G_{N}\mathcal{T}^{\mu\nu}$ as the variation of the metric directly from the Lagrangian (\ref{eq:defom:action:generic:int}).
 With these simple heuristic arguments we have shown that field theory in compact
space-time is compatible with GR. 

In forthcoming papers\footnote{One of these papers is now published in \emph{Annals of Physics},  \cite{Dolce:tune}.},  we will show that, by writing (\ref{eq:deform:4mom:generic:int})
as a minimal substitution, such a description of interactions
can also be used to describe ordinary gauge interactions in terms of space-time geometrodynamics. Gauge fields
turn out to encode  modulations of periodicities associated to local transformations of reference frame.  
The assumption of  PBCs at the geometrodynamical boundary  of the theory  lead, through a generalization of the demonstration of a generalization of (\ref{periodic:path.integr:Oper:Fey}), to the ordinary FPI of scalar QED, \cite{Dolce:tune}. Such a semi-classical description of QED yield, for instance, a intuitive description of superconductivity \cite{Dolce:2009cev4,Dolce:tune,Dolce:SuperC}. 

We have also mentioned that a cyclic field turns out to be dual an XD field,  \cite{Dolce:2009cev4,Dolce:AdSCFT}. In fact  the compact world-line
parameter of the theory plays the role of a VXD. On the other hand we have shown
that a cyclic field reproduces ordinary quantum dynamics \cite{Dolce:2009ce,Dolce:tune}.
From the dualism to XD theories and from the geometrodynamical approach
to interactions described above, we find that
the classical evolution of the periodic fields along a deformed VXD
background reproduces  the quantum behavior of the corresponding
interaction scheme. Thus  relativistic field theory in compact 4D provides the possibility
of an intuitive, unconventional interpretation of the Maldacena conjecture. In fact, according to  Witten, in AdS/CFT ``quantum
phenomena [...] are encoded in classical geometry''. We will
apply this idea to a simple Bjorken Hydrodynamical Model for Quark-Gluon-Plasma
(QGP) logaritmic freeze-out \cite{Magas:2003yp}. In first approximation
the energy momentum of the QGP can be supposed to decay exponentially
(similarly to the Newton's law of cooling for a thermodynamic system \cite{Satz:2008kb}). This interaction scheme is therefore described by the conformal warped tetrad $e_{\mu}^{a}=\delta_{\mu}^{a}e^{-ks}$,
where $s$ is the proper time, \emph{i.e.} by a \emph{virtual} AdS metric.
As a result, the classical configurations of cyclic fields in such a deformed
background reproduces basic, phenomenological aspects of AdS/QCD, \cite{Dolce:AdSCFT}.

\section{Conclusions}

The formalism of field theory in  compact space-time dimensions  provides, through PBCs, a fully consistent, natural description of both the digital aspect of reality arising QM and the analog aspect of reality typical of relativity \cite{Dolce:2009ce,Dolce:tune,Dolce:AdSCFT,Dolce:2009cev4,Dolce:Dice,Dolce:2010ij,Dolce:2010zz,Dolce:cyclic,Dolce:SuperC}. It must be noticed that (general and special) relativity sets the differential structure of space-time without giving any particular prescription for the BCs. On the other hand, the BCs have played an important role since the earliest days of QM (for instance as in the Bohr atom or as in the particle in a box). We have seen that relativity is compatible with compact (analog) space-time dimensions, as long as the periodicities (\emph{i.e.} the compactification lengths) are allowed to transform in a covariant way. On the other hand, through discrete Fourier transform to a compact dimension there is naturally associated a discretized frequency spectrum, and thus a quantized energy spectrum. 

The physical assumption of intrinsic periodicity leads to a formal correspondence with ordinary (axiomatic) relativistic QM in both the canonical and the Feynman formulations, as well as for many non trivial quantum phenomena, \cite{Dolce:2009ce,Dolce:tune,Dolce:AdSCFT,Dolce:2009cev4,Dolce:Dice,Dolce:2010ij,Dolce:2010zz,Dolce:cyclic,Dolce:SuperC}. Time periodicity can be used to describe naturally the transition between the classical  and the quantum regime of an elementary systems. We may consider for instance, the different components of an electromagnetic field in a Black-Body radiation, \emph{i.e.} in the case of massless cyclic phenomena. The IR components correspond to the limit of nearly infinite time periodicity, \emph{i.e.} low energy or frequency. In this case the PBCs can be neglected and the elementary system is described by fields with approximatively continuous energy spectrum (in this purely analog limit the thermal noise destroys the intrinsic periodicity in a sort of decoherence). In the limit of small time periodicity, corresponding to the UV components of a Black Body radiation, the field description of the theory is gradually replaced by digital corpuscular aspects. The PBCs can not be neglected and the energy spectrum turns out to be quantized (the thermal noise is not sufficient to destroy the intrinsic periodicity).  The initial and final configurations of a periodic field form a periodic space-time lattice, so that the cyclic evolution is described by a sum over classical paths with different winding numbers (an electron does $10^10$ cycles for every ``tick'' of the Cs clock).   

These intrinsic time periodicities of elementary particles can be identified with the so call ``de Broglie internal clocks'' or ``de Broglie periodic phenomena'' at the base of the undulatory description of modern relativistic QM. Similarly to a stopwatch, every moment in time is determined by the combination of the phases or the ``ticks'' of  periodic cycles (\emph{e.g.} years, months, days, hours, minutes and seconds). That is, every value of our external temporal axes (defined with reference to the digital ``ticks'' of the Cs-133 atomic clock) is characterized by a unique combination of the ``ticks'' of all the ``de Broglie internal clocks'', \emph{i.e.} elementary particles, constituting the system under investigation. In this scenario the local nature of relativistic time turns out to be enforced. The long temporal (and spatial) scales are provided by massless fields with low frequencies (the long compactifications lengths provides the underlying spatial and temporal structure of the system of elementary particles). As we have seen, it must be considered that the such clocks can modulate periodicity through interactions (exchange of energy) and that the periodicities depend dynamically on reference systems according to the relativistic laws (\emph{e.g.} relativistic Doppler effect). This also means that the combination of two or more de Broglie clocks, \emph{i.e.} a non elementary system of particles, leads to  very  chaotic evolutions as interaction is turned on.  Such a description of the flow of time depends only on the reciprocal combinations and variations of the ``ticks'' of the de Broglie clocks. It is therefore reference frame depended in a relativistic way. Moreover the arms of the de Broglie internal clocks can be conventionally supposed to rotate clockwise or anticlockwise (but an inversion of a single clock corresponds to transform the particle to the corresponding antiparticle or \emph{vice versa}).  These further conceptual aspects are particularly interesting for the problem of the time arrow in physics.
\pagebreak
 \bibliographystyle{aipproc}

\begin{thebibliography}{23}
\expandafter\ifx\csname natexlab\endcsname\relax\def\natexlab#1{#1}\fi
\providecommand{\enquote}[1]{``#1''}
\expandafter\ifx\csname url\endcsname\relax
  \def\url#1{\texttt{#1}}\fi
\expandafter\ifx\csname urlprefix\endcsname\relax\def\urlprefix{URL }\fi
\providecommand{\eprint}[2][]{\url{#2}}

\bibitem{Dolce:2009ce}
Dolce D 2011 {\em Found. Phys.\/} {\bf 41} 178,  
\textit{Preprint} \eprint{0903.3680v5}

\bibitem{Dolce:tune}
Dolce D  2012 {\em Ann. Phys.\/} {\bf 327} (6) 1562,  Gauge interaction as periodicity modulation. \textit{Preprint}
  \eprint{1110.0315}

\bibitem{Dolce:AdSCFT}
Dolce D 2011 AdS/CFT in Virtual Extra Dimension  \textit{Preprint}
  \eprint{1110.0316}

\bibitem{Dolce:2009cev4}
Dolce D 2009 Compact Time and Determinism for Bosons \textit{Preprint} \eprint{0903.3680v1-v4} (\emph{partially published in} [1])

\bibitem{Dolce:cyclic}
Dolce D 2012 {\em J. phys.: Conf. Ser.\/} {\bf 343} 012031

\bibitem{Dolce:Dice}
Dolce D 2011 {\em J. phys.: Conf. Ser.\/} {\bf 306} 10
\textit{Preprint} \eprint{1111.3319}

\bibitem{Dolce:2010ij}
Dolce D 2010 {\em AIP Conf. Proc.\/} {\bf 1246} 178 
\textit{Preprint}  \eprint{1006.5648}

\bibitem{Dolce:2010zz}
Dolce D 2010 {\em AIP Conf. Proc.\/} {\bf 1232} 222 
\textit{Preprint}  \eprint{1001.2718}

\bibitem{Dolce:SuperC}
Dolce D 2012 Gauge symmetry breaking without the \textit{v.e.v.} and other considerations about superconductivity {(\em Draft)\/} 


\bibitem['t~Hooft(2003)]{'tHooft:2001ar}
G.~'t~Hooft 2003 \emph{Int. J. Theor. Phys.} \textbf{42}, 355,
 \textit{Preprint} \eprint{hep-th/0104080}.


\bibitem['t~Hooft(2001)]{'tHooft:2001fb}
G.~'t~Hooft  2001 Quantum Mechanics and Determinism \textit{Preprint} \eprint{hep-th/0104219}.

\bibitem[Elze and Schipper(2002)]{Elze:2002eg}
H.-T. Elze, and O.~Schipper, 2002 \emph{Phys. Rev.} \textbf{D66} 044020
  \textit{Preprint} \eprint{gr-qc/0205071}.

\bibitem['t~Hooft(2007{\natexlab{a}})]{'tHooft:2006sy}
G.~'t~Hooft 2007 \emph{J. Phys.: Conf. Ser.} \textbf{67} 15 \textit{Preprint} \eprint{quant-ph/0604008}.

\bibitem['t~Hooft(2007{\natexlab{b}})]{'tHooft:2007xi}
G.~'t~Hooft 2007 \emph{AIP Conf. Proc.} \textbf{957} 154 \textit{Preprint} \eprint{0707.4568}.

\bibitem[Jaffe(2005)]{Jaffe:2005vp}
R.~L. Jaffe 2005 \emph{Phys. Rev.} \textbf{D72}, 021301 
 \textit{Preprint} \eprint{hep-th/0503158}.

\bibitem['t~Hooft(2010)]{'tHooft:2010zzb}
G.~'t~Hooft 2010 \emph{Int. J. Mod. Phys.} \textbf{A25}, 4385.

\bibitem[Einstein(1910)]{Einstein:1910}
A.~Einstein 1910 \emph{Arch. Sci. Phys. Natur.} \textbf{29}.

\bibitem[{Ferber}(1996)]{1996FoPhL}
R.~{Ferber} 1996 \emph{Found. Phys. Lett.} \textbf{9}  6
  575.

\bibitem[{Catillon} et~al.(2008)]{2008FoPh...38..659C}
P.~{Catillon}, N.~{Cue}, M.~J. {Gaillard}, R.~{Genre}, M.~{Gouan{\`e}re}, R.~G.
  {Kirsch}, J.-C. {Poizat}, J.~{Remillieux}, L.~{Roussel}, and M.~{Spighel}, 2008
  \emph{Found. Phys.} \textbf{38} 659.

\bibitem[Matsubara(1955)]{Matsubara:1955ws}
T.~Matsubara 1955 \emph{Prog. Theor. Phys.} \textbf{14} 351.

\bibitem[Kaluza(1921)]{Kaluza:1921tu}
T.~Kaluza 1921 \emph{Sitzungsber. Preuss. Akad. Wiss. Berlin (Math. Phys. )}
  \textbf{1921}, 966.

\bibitem[Feynman(1942)]{Feynman:1942us}
R.~P. Feynman 1942 The Principle of Least Action in Quantum Mechanics \emph{PhD Thesis}.

\bibitem[Nikolic(2007)]{Nikolic:2006az}
H.~Nikolic 2007 \emph{Found. Phys.} \textbf{37}, 1563--1611 (2007),
\textit{Preprint}  \eprint{quant-ph/0609163}.

\bibitem[Ohanian and Ruffini(????)]{Ohanian:1995uu}
H.~Ohanian, and R.~Ruffini 1994 New York, USA: Norton, 679.

\bibitem[York(1986)]{springerlink:10.1007/BF01889475}
J.~York 1986 \emph{Found. Phys.} \textbf{16} 249

\bibitem[Magas et~al.(2003)]{Magas:2003yp}
V.~K. e.~a. Magas, A.~Anderlik, C.~Anderlik, and L.~P. Csernai, 2003 \emph{Eur.
  Phys. J.} \textbf{C30} 255 \textit{Preprint} \eprint{nucl-th/0307017}.

\bibitem[Satz(2008)]{Satz:2008kb}
H.~Satz  2008 The Thermodynamics of Quarks and Gluons \textit{Preprint} \eprint{0803.1611}.

\end{thebibliography}

\end{document}